\documentclass[pdflatex,sn-mathphys-num]{sn-jnl}% Math and Physical Sciences Numbered Reference Style
\usepackage{graphicx}%
\usepackage{multirow}%
\usepackage{amsmath,amssymb,amsfonts}%
\usepackage{amsthm}%
\usepackage{mathrsfs}%
\usepackage[title]{appendix}%
\usepackage{xcolor}%
\usepackage{textcomp}%
\usepackage{manyfoot}%
\usepackage{booktabs}%
\usepackage{algorithm}%
\usepackage{algorithmicx}%
\usepackage{algpseudocode}%
\usepackage{listings}%
\usepackage{mathtools}%
\theoremstyle{thmstyleone}%
\theoremstyle{thmstyletwo}%

\theoremstyle{thmstylethree}%

\begin{document}

\title[Molecular Design beyond Training Data with Novel Extended Objective Functionals of Generative AI Models Driven by Quantum Annealing Computer]{Molecular Design beyond Training Data with Novel Extended Objective Functionals of Generative AI Models Driven by Quantum Annealing Computer}


\author[1]{\fnm{Hayato} \sur{Kunugi}}
\author[2]{\fnm{Mohsen} \sur{Rahmani}}
\author[1]{\fnm{Yosuke} \sur{Iyama}}
\author[1]{\fnm{Yutaro} \sur{Hirono}}
\author[1]{\fnm{Akira} \sur{Suma}}
\author[2]{\fnm{Matthew} \sur{Woolay}}
\author[2]{\fnm{Vladimir} \sur{Vargas-Calder\'{o}n}}
\author[2]{\fnm{William} \sur{Kim}}
\author[2]{\fnm{Kevin} \sur{Chern}}
\author[2]{\fnm{Mohammad} \sur{Amin}}
\author*[1]{\fnm{Masaru} \sur{Tateno}}\email{tateno1611@gmail.com}

\affil[1]{\orgdiv{Inovation to Implementation Laboratories, Central Pharmaceutical Research Institute}, \orgname{Japan Tobacco Inc.}, \orgaddress{\street{1-1 Murasaki-cho}, \city{Takatsuki}, \postcode{569-1125}, \country{Japan}}}
\affil[2]{\orgname{D-Wave Systems Inc.}, \orgaddress{\street{3033 Beta Ave}, \city{Burnaby}, \postcode{V5G 4M9}, \country{Canada}}}


\abstract{Deep generative modeling to stochastically design small
molecules is an emerging technology for accelerating drug discovery and
development. However, one major issue in molecular generative models is
their lower frequency of drug-like compounds. To solve this problem, we
develop a novel, quantum annealing, generative-model approach for
optimizing deep generative models, an approach that includes integrating 
with a D-Wave annealing quantum computer. Of particular note, our neural 
hash function (NHF) is used simultaneously as a regularization scheme
and binarization scheme, of which the latter is for transformation
between continuous and discrete signals of the classical and quantum
neural networks, respectively, in the error evaluation (i.e., objective)
function. The compounds generated via the quantum-annealing generative
models exhibit higher quality in both validity and drug-likeness than
those generated via the fully-classical models, and even exceed the
training data in terms of drug-likeness features, without any restraints
and conditions to deliberately induce such an optimization. These
results suggest that quantum annealing could be used as a stochastic
generator integrated with our novel neural network architectures for
extending the performance of feature space sampling and the extraction
of characteristic features in drug design.}

\maketitle

\section{Introduction}\label{introduction}

In the field of drug discovery, efficiently designing molecular
structures with optimal chemical properties and synthetic accessibility
is a complex and important research area. Typical approaches such as
iterating through the design and experiment cycle can only search a
small region in a vast chemical space, where the number of synthesizable
molecules of size acceptable for drugs is estimated to be over
10\textsuperscript{60} \cite{bohacek_art_1996}. Machine learning and deep
learning-based drug design can efficiently explore a wider region in
such a huge chemical space. Actually, by applying recent achievements in
deep learning and deep generative models, various deep generative models
have been reported for molecules with desired chemical properties
\cite{ajagekar_molecular_2023, dollar_attention-based_2021, zhao_comprehensive_2025}.

Despite these efforts, two challenges with compounds generated from
existing generative models remain: 1) the low frequency of ``drug-like''
molecules that satisfy the activity to the target proteins and have
acceptable chemical properties and synthetic accessibility; and 2) a
trade-off between the character of the compounds and the diversity in
the chemical structure \cite{lambrinidis_challenges_2018}. One cause of
these problems is the lack of data, resulting in low generalization due
to overfitting. Currently, the number of compounds that can be
synthesized is \ensuremath{\sim} 10\textsuperscript{10} \cite{sadybekov_synthon-based_2022},
whereas the expected size of the chemical space is 10\textsuperscript{60} (as
previously mentioned)---thus, the number of synthesizable compounds
represents only a very small proportion of the entire chemical search
space, and is insufficient for use as training data.

Quantum machine learning (QML) (also referred to as quantum artificial
intelligence (QAI)) is a growing area of research that combines quantum
computing and machine learning. QML investigates how quantum resources
such as superposition, entanglement, and tunneling can accelerate or
enhance classical learning models. Most early work has focused on the
gate-based paradigm, where data and model parameters are encoded into
quantum circuits composed of unitary transformations and projective
measurements. Algorithms such as quantum support vector machines, kernel
estimators, and quantum neural networks are implemented through
parameterized quantum circuits, optimized using hybrid quantum-classical
feedback loops \cite{devadas_quantum_2025}. However, in many common
scenarios, training parameterized quantum circuits has been shown to be
unlikely to be able to scale due to trainability issues known as barren
plateaus \cite{mcclean_barren_2018}; it may also be possible for
classical computing to efficiently simulate such quantum circuits
\cite{cerezo_does_2025}.

An alternative and more physically motivated framework arises in quantum
annealing \cite{albash_adiabatic_2018, johnson_quantum_2011}, which
provides an analog realization of optimization and sampling tasks
central to machine learning. In quantum annealing, learning problems are
mapped onto an Ising Hamiltonian and the system searches for the
low-energy configurations. Sampling from these configurations can be
guided by training the parameters of the Ising Hamiltonian to learn a
``binary'' data distribution or a ``binary'' latent representation of an
autoencoder. Quantum annealers, such as D-Wave's Advantage2\textsuperscript{TM} quantum
computer, implement these dynamics in hardware, enabling large-scale
exploration of complex energy landscapes for machine learning
applications
\cite{amin_quantum_2018, benedetti_quantum-assisted_2017, winci_path_2020}.
Recent experiments have demonstrated that quantum annealing achieves a
scaling advantage in approaching low-energy configurations
\cite{king_quantum_2023, king_beyond-classical_2025}, while the
resulting sampling distributions cannot be efficiently reproduced by any
known classical simulation \cite{king_beyond-classical_2025}.
Therefore, it is important to explore ways to harness this intrinsic
sampling power for quantum-enhanced generative modeling in areas such as
drug discovery and materials design.

In this work, our approach started from Variational Autoencoder
(VAE)-based generative models for generation and inference of chemical
compounds. VAE sets the approximated posterior distribution of the
latent variables and optimizes the evidence lower bound (ELBO) instead
of the true log-likelihood, which is generally intractable. By the
amortized inference and the reparameterization framework
\cite{kingma_auto-encoding_2013, rezende_stochastic_2014}, VAE can
efficiently train its objective function, which is also referred to as
the loss function with the reconstruction loss and regularization terms
included, and is widely used for the generative model.
TransVAE \cite{dollar_attention-based_2021} also involves the VAE that
generates character sequences of molecules via a combination of
Transformer-based Encoder/Decoder and continuous-valued latent space.
More recently, VAE with discrete latent variables, Discrete VAE (DVAE)
\cite{rolfe_discrete_2016}, in which the generative process is driven
by a Boltzmann Machine (BM), was adopted for the generation of chemical
compounds \cite{gircha_hybrid_2023}.

DVAE incorporates discrete latent variables with a discrete prior,
making it a more suitable alternative for modeling data with categorical
structures such as molecular descriptors tokenized by employing a
particular transformation scheme. In addition, DVAE is suitable for
combining a VAE architecture with quantum computing in which measurement
outcomes are binary (spin) states. Quantum VAE
\cite{khoshaman_quantum_2018} replaced the prior distribution from a
classical BM with a quantum Boltzmann machine (QBM)
\cite{amin_quantum_2018} in which the D-Wave quantum annealer leveraged
the sampling from the Boltzmann distribution.

In DVAE, converting a continuous representation of the input data to a
binary representation in the final encoding step, is not differentiable
and, thus, prevents gradients of the error (reconstruction loss) from
being backpropagated to the encoder. To solve this issue, DVAE
introduced auxiliary continuous variables for training the approximating
posterior, which stochastically converted from the discrete variables
\cite{rolfe_discrete_2016, khoshaman_gumbolt_2018}. Considering a
specific form of the smoothing distribution, the error can be propagated
back through the latent variables by reparameterization. During
training, only discrete variables are used in the prior distribution,
whereas continuous variables are used in the approximating posterior.
Thus, non-differentiability for backpropagation is still a crucial issue
in this scheme.

We propose another approach to solve the non-differentiability of the
encoder by involving our novel scheme in the objective function of the
autoencoder-based generative model. In this scheme, inspired by Deep
Hashing \cite{erin_liong_deep_2015}, the output of the encoder is
binarized using outputs of a ``deterministic'' function (\emph{i.e.}, a
stochastic distribution is not involved) as the latent variables. An
additional term that aims to reduce the loss in binarization is included
in the total objective (loss) function for error evaluation of the
system. Carefully defining the structure of the loss function and its
derivatives enables the loss to backpropagate through the latent
variables without smoothing and stochastic reparameterization
(differentiability). We applied this scheme to our generative models of
chemical compounds, thereby showing that employment of this scheme
improved the validity and drug-likeness of the generated compounds in
both classical and quantum computations of the prior distributions.

In addition, we utilized QBM as a prior distribution of the compound
generation model, thus indicating that the quantum prior outperforms the
classical counterpart in the quality of the generated compounds.
Interestingly, the compounds generated via our ``quantum'' generative
model exhibited a higher quality in both structural validity and
drug-likeness scores than those of even the training dataset without any
conditions to deliberately induce optimal trends of the evaluation
scores. Thus, our present analysis suggests that our novel objective
functionals integrated with the D-Wave quantum annealer should possess a
powerful potential for the sampling task of the drug discovery field.

\bigskip

\section{Results}\label{results}

\subsection{Quantum Annealing}\label{quantum-annealing}

To solve the learning problems using quantum annealing, the Ising
Hamiltonian is formulated as
\begin{equation}
\begin{split}
H(s) &= A(s) H_D + B(s) H_P
\\
H_D &= -\sum_i
\sigma_{i}^x,
\hspace{0.3in}
H_P = \sum_{i, j}
J_{ij} \sigma_i^z \sigma_j^z +
\sum_i h_i \sigma_i^z
\\
\end{split} \label{eq-qa-hamiltonian}
\end{equation}
where \(H_D\) is a transverse-fiels driver promoting quantum tunneling,
\(H_P\) encodes the cost function or model parameters and
\(\sigma_i^x, \sigma_i^z\) are Pauli operators for \(i\)-th element. By
slowly varying the annealing schedule \(A(s)\) and \(B(s)\), the system
evolves toward the low-energy configurations of \(H_P\). Sampling from
these configurations can be guided by training the parameters \(J_{ij}\)
and \(h_i\) to learn the prior distribution of binary latent variables
(see Equation~\ref{eq-elbo}).

\subsection{VAE and DVAE}\label{vae-and-dvae}

The training framework of the original VAE is formulated by the
variational inference, which maximizes ELBO on the true log-likelihood
of data:
\begin{equation}
\mathcal{L} =
\mathbb{E}_{q_\phi(\mathbf{z}|\mathbf{x})}
[\log p_\theta(\mathbf{x} | \mathbf{z})] -
D_\mathrm{KL} \left(
    q_\phi(\mathbf{z} | \mathbf{x})
    \ \middle\| \ 
    p_\psi(\mathbf{z})
\right)
\leqq
\log p_\theta (\mathbf{x}),
\label{eq-elbo}
\end{equation}
where \(\phi, \theta\) and \(\psi\) denote the trainable parameters for
the encoder, decoder and prior. The prior parameters \(\psi\) includes
the parameters of Ising Hamiltonian \(J_{ij}, h_i\) (see
Equation~\ref{eq-qa-hamiltonian}). The first term corresponds to the
reconstruction loss which measures the expected log-likelihood (decoder)
\(p_\theta(\mathbf{x} | \mathbf{z})\) of the data \(\mathbf{x}\) given
the latent variables \(\mathbf{z}\) under the approximate posterior
(encoder) \(q_\phi(\mathbf{z} | \mathbf{x})\). The second term
represents the Kullback-Liebler (KL)-divergence between the approximated
posterior \(q_\phi(\mathbf{z} | \mathbf{x})\) and the prior
\(p_\psi(\mathbf{z})\). The training objective is
\begin{gather}
\underset{\phi, \theta, \psi}{\mathrm{argmin}}
\ \ 
\mathbb{E}_{\mathbf{x} \sim \mathcal{D}}
[L_\mathrm{elbo}(\mathbf{x}; \phi, \theta, \psi)]
\label{eq-train-obj} \\
L_\mathrm{elbo} \coloneqq
\underbrace{
    -\mathbb{E}_{q_\phi(\mathbf{z} | \mathbf{x})}
    [\log p_\theta(\mathbf{x} | \mathbf{z})]
}_{
    L_\mathrm{rec}(\mathbf{x}; \phi, \theta)
}
+ D_\mathrm{KL}
\left(
    q_\phi(\mathbf{z} | \mathbf{x})
    \ \middle\| \ 
    p_\psi(\mathbf{z})
\right)
\label{eq-elbo-decomposed}
\end{gather}
where \(\mathcal{D}\) denotes the data (empirical) distribution.

DVAE \cite{rolfe_discrete_2016} incorporates discrete latent variables
with a discrete prior, making DVAE a better alternative for modeling
data with categorical structures. Simplified Molecular Input Line Entry
System (SMILES) \cite{weininger_smiles_1988} and SELF-referencing
Embedded Strings (SELFIES) \cite{krenn_selfies_2022} are widely used
for encoding molecules to sequences of strings. We created molecular
descriptors tokenized from SMILES strings. In addition, DVAE is suitable
for combining the VAE architecture with quantum computing where the
observed values are binary (spin) states. DVAE proposed a stochastic
binarization to characterize the posterior distribution
\(q_\phi(\mathbf{z} | \mathbf{x})\) for \(D\)-dimensional discrete
latent variables \(\mathbf{z} = (z_i)_{i=1, \cdots, D} \in \{0, 1\}^D\)
from outputs of the encoder \(\mathbf{l} = (l_i)_{i=1, \cdots, D}\) as
follows:
\begin{equation}
\begin{split}
q_i &= \sigma(l_i)
\\
z_i &= \mathcal{H}(q_i + \rho_i - 1)
= \mathcal{H} \left(
    l_i + \sigma^{-1}(\rho_i)
\right),
\hspace{0.2in}
\rho_i \sim \mathrm{Unif}(0, 1),
\\
\end{split} \label{eq-gumbel}
\end{equation}
where
\(\sigma(x) = \frac{1}{1 + e^{-x}}, \sigma^{-1} (y) = \log y - \log (1 - y)\)
denote the sigmoid function as its inverse, and \(\mathcal{H}\) denotes
the Heaviside step function. The main obstacle of VAE with discrete
latent variables is that the function \(z = z_\phi(x, \rho)\) is
non-differentiable such as Heaviside, which prevents backpropagation
using the reparameterization trick. Gumbolt
\cite{winci_path_2020, khoshaman_gumbolt_2018} applied the Gumbel trick
\cite{maddison_concrete_2016}: a smoothing of \(\mathbf{z}\) to obtain
a continuous variable
\(\boldsymbol{\zeta} = (\zeta_i)_{i=1, \cdots, D} \in (0, 1)^D\) by
\(\zeta_i = \sigma \left( \frac{l_i + \sigma^{-1} (\rho_i)}{\tau} \right)\),
where \(\tau\) is a smoothing parameter with
\(\boldsymbol{\zeta} \to \mathbf{z}\) in the limit of \(\tau \to 0\).

\subsection{Involvement of the Neural Hash Function
(NHF)}\label{involvement-of-the-neural-hash-function-nhf}

Herein, we aimed to solve the afore-mentioned, non-differentiable
features derived from the binarization of discrete latent variables. We
first constructed a Transformer \cite{vaswani_attention_2017}-based
encoder-decoder architecture with discrete latent variables.

Let \(\mathbf{X} \in \mathbb{N}^{N \times C}\) represent the tokenized
SMILES of \(N\) molecules using a vocaburary size of \(V\), where \(C\)
denotes the maximum number of tokens, Each element
(\(X_{nc} \in \{1, 2, \cdots, V\}\)) of \(\mathbf{X}\) represents the
index of the \(c\)-th token of the \(n\)-th molecule. Tokens are
embedded in a \(d_\mathrm{model}\)-dimensional space and fed into an
encoder \(f_\phi\) including Transformer layers, a Neural Tensor Network
block, and MLP layers (see Methods) to obtain the \(D\)-dimensional
fixed length vectors
\(\mathbf{H} = (\mathbf{h}_1, \mathbf{h}_2, \cdots, \mathbf{h}_N) \in \mathbb{R}^{N \times D}\),
Binary latent variables
\(\mathbf{Z} = (\mathbf{z}_1, \mathbf{z}_2, \cdots, \mathbf{z}_N) \in \{0, 1\}^{N \times D}\)
are obtained from \(\mathbf{H}\) through a binarization function and fed
into a decoder \(g_\theta\) including another Neural Tensor Network and
Transformer layers. Finally, outputs of the decoder are passed through a
softmax layer to get probabilities of reconstructed tokens
\(\hat{\mathbf{X}} \in \mathbb{R}^{N \times C \times V}\).

Inspired by Deep Hashing \cite{erin_liong_deep_2015}, we derived the
loss function \(L_\mathrm{nhf}\) from the ELBO of the joint distribution
of the binary latent variables \(\mathbf{z}\) and the continuous ones
\(\mathbf{h}\) (for details, see the Discussion section), referred to
here as the neural hash function (NHF), as the objective (loss) function
of our generative models (see Methods for details):
\begin{align}
L_\mathrm{nhf} &=
L_\mathrm{rec}(\mathbf{X}; \theta, \phi) +
L_\mathrm{prior}(\mathbf{Z}; \phi, \psi) +
L_\mathrm{quant}(\mathbf{Z}; \phi)
\label{eq-loss-nhf} \\
L_\mathrm{rec}(\mathbf{X}; \theta, \phi)
&\coloneqq
\frac{1}{N}
\sum_{n=1}^N
\mathrm{CE}(\mathbf{X}_n, \hat{\mathbf{X}}_n)
\label{eq-loss-rec} \\
L_\mathrm{prior}(\mathbf{Z}; \phi, \psi)
&\coloneqq
-\frac{1}{N}
\log p_\psi(\mathbf{z}_n)
\label{eq-loss-prior} \\
L_\mathrm{quant}(\mathbf{Z}; \phi)
&\coloneqq
\frac{\lambda_\mathrm{fro}}{2N}
\|\mathbf{Z} - \mathbf{H}\|_F^2 +
\frac{\lambda_\mathrm{ortho}}{2}
\sum_{l=1}^L
\| \mathbf{W}_l \mathbf{W}_l^T - \mathbf{I}\|_F^2,
\label{eq-loss-quant}
\end{align}
where \(\mathrm{CE}\) denotes the cross entropy and \(\|\cdot\|_F\)
denotes the Frobenius norm (\(L_2\) norm) defined for matrices.
\(L_\mathrm{rec}\) is the reconstruction loss as the cross entropy
between inputs and reconstructed tokens. \(L_\mathrm{prior}\) is the
approximated cross entropy of the prior distribution
\(p_\psi(\mathbf{z})\) and the empirical distribution.
\(L_\mathrm{quant}\) means the quantization loss to produce good binary
codes. The second term encourages the weight matrices
\(\mathbf{W}_l, l=1,2,\cdots, L\) of the \(L\)-layer MLP in the encoder
to be orthogonal, that is, requiring the different dimensions to be
independent.

In Deep Hashing, the transformation from \(\mathbf{H}\) to
\(\mathbf{Z}\) is performed by a non-smooth function such as Heaviside
function. To optimize the model parameters by the stochastic gradient
descent method, we defined the gradient of \(L_\mathrm{nhf}\) with
respect to the parameters of the decoder \(\theta\), prior \(\psi\) and
encoder \(\phi\) as follows (see Methods for details):
\begin{align}
\frac{
    \partial L_\mathrm{nhf}
}{
    \partial \theta
}
&= \frac{
    \partial
    L_\mathrm{rec}(\mathbf{X}; \theta, \phi)
}{
    \partial \theta
}
\label{eq-deriv-nhf-theta} \\
\frac{
    \partial L_\mathrm{nhf}
}{
    \partial \psi
}
&= \frac{
    \partial
    L_\mathrm{prior}(\mathbf{Z}; \phi, \psi)
}{
    \partial \psi
}
\label{eq-deriv-nhf-psi} \\
\frac{
    \partial L_\mathrm{nhf}
}{
    \partial \phi
}
&=
\sum_{n=1}^N
\boldsymbol{\delta}_n^T
\frac{\partial \mathbf{h}_n}{\partial \phi}
\label{eq-deriv-nhf-phi} \\
\boldsymbol{\delta}_n
&\coloneqq
\frac{\lambda_\mathrm{fro}}{N}
(\mathbf{z}_n - \mathbf{h}_n) +
\frac{
    \partial
    L_\mathrm{rec}(\mathbf{X}; \theta, \phi)
}{
    \partial \mathbf{z}_n
} +
\frac{
    \partial
    L_\mathrm{prior}(\mathbf{Z}; \phi, \psi)
}{
    \partial \mathbf{z}_n
} \label{eq-define-delta}
\end{align}

\begin{figure}[ht]
\centering
\includegraphics[width=\linewidth]{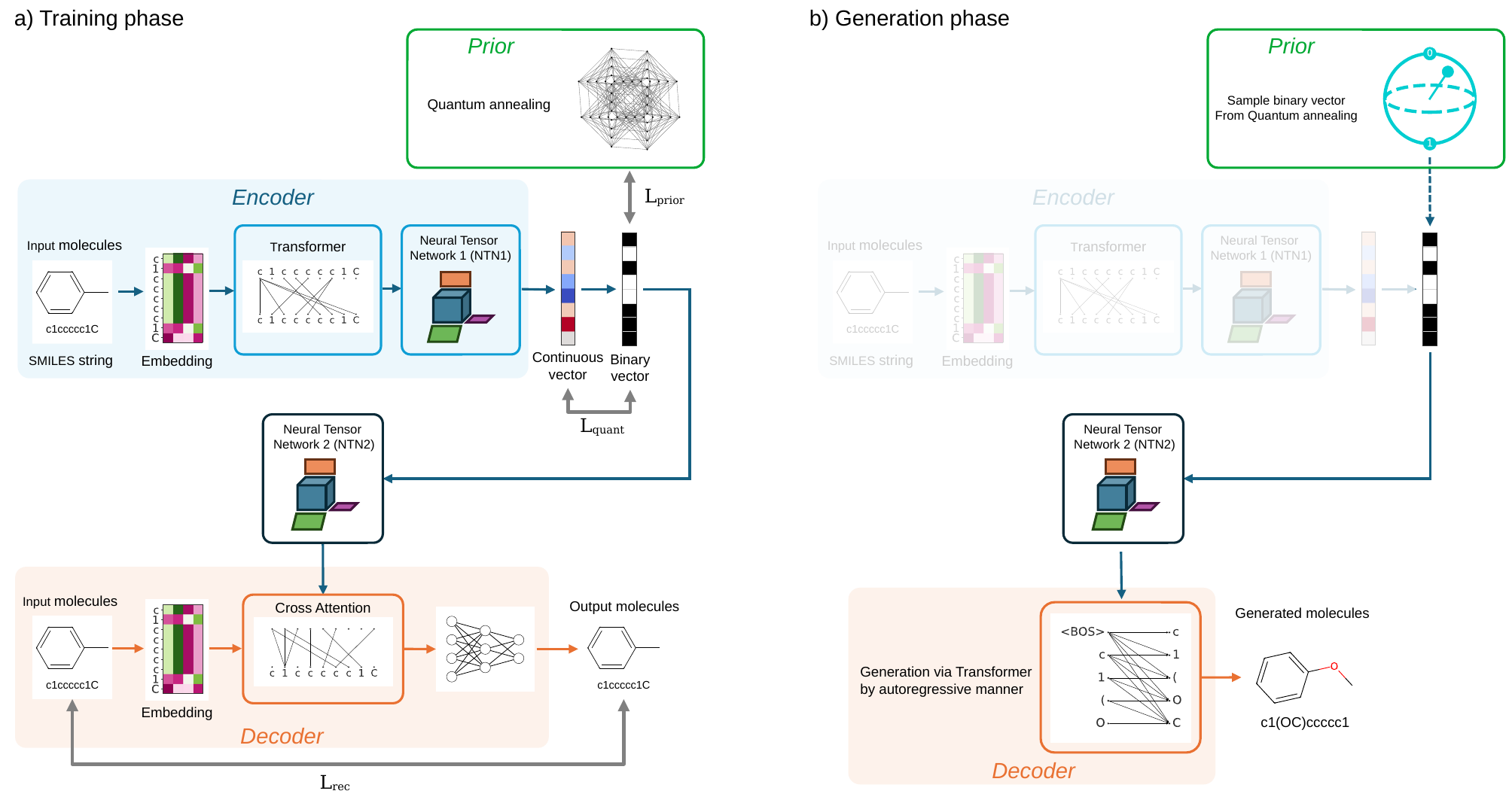}
\caption{\textbf{Scheme of the Neural Hash Function
(NHF)-based generative model.} (a) Input compound \(\mathbf{X}\) is
represented by employing SMILES sequences. Each token from SMILES is
embedded in a 160-dimensional vector and added the positional encoding.
The embedded tensor is fed into the Transformer encoder layers and
flattened by the Neural Tensor Network layer (NTN1) to obtain the
continuous vector. \(\mathbf{h}\) is transformed by our NHF to the
binary vector, and then transformed into the latent matrix through
another NTN layer (NTN2). In decoder block, the input tensors are passed
through the self-attention layer with the subsequent masks. The
cross-attention between the output of the self-attention layer and the
latent matrix. Finally, SMILES sequences are reconstructed from outputs
of the decoder through the softmax layer. The reconstruction loss
between the input and reconstructed tokens (\(L_\mathrm{rec}\)), the
cross entropy of the prior distribution (\(L_\mathrm{prior}\)) and the
quantization loss to produce good binary codes (\(L_\mathrm{quant}\))
are computed for updating of parameters of the encoder, decoder and
prior (see Equation~\ref{eq-loss-nhf}). (b) In the generation phase, the
binary latent variables are sampled from the Quantum or classical
Boltzmann Machine (BM) prior and transformed into the latent matrix
through the NTN2. The output SMILES sequences are predicted by
autoregressive manner.}\label{fig-scheme}
\end{figure}

The scheme of the entire model and the workflow of the training and
generation phase are shown in Figure~\ref{fig-scheme}.

We examined the effect of the molecular generation model using the NHF.
This model inputs and outputs the SMILES sequences as representations of
chemical structures. We trained the model using the ChEMBL public
dataset, and generated the compounds from the BM prior and the Decoder.
We evaluated the validity of the generated SMILES, a fraction of SMILES
strings that were successfully converted into molecular structures, as a
metric for generative models. Our NHF outperformed the validity of the
Gumbel-Softmax binarization (52.2\% in Gumbel-Softmax and 62.0\% in NHF
for the fully classical computations) (Table~\ref{tbl-metrics}).

\begin{table}[h]
\caption{Metrics of compounds generated by Transformer-based encoder-decoder models.}\label{tbl-metrics}
\begin{tabular}{@{}ccccc@{}}
\toprule
Sampler & Binarization & Validity (\%) & Uniqueness (\%) &
\begin{tabular}{c}
Unique compounds / \\ generated compounds
\end{tabular}
\\
\midrule
Classical BM & Gumbel-Softmax & 52.20                      & \underline{\textbf{99.94}} & 5126 / 10000 \\
             & NHF            & 61.95                      & \textbf{98.09}             & \textbf{6077 / 10000} \\
QBM          & Gumbel-Softmax & \textbf{71.88}             & 95.10                      & \underline{\textbf{6835 / 10000}} \\
             & NHF            & \underline{\textbf{96.97}} & 51.92                      & 5035 / 10000 \\
\botrule
\end{tabular}
\footnotetext{
    Validity is a fraction of the SMILES sequences that can be interpreted to the molecular graph by employing RDKit. Uniqueness is a fraction of compounds that are uniquely (not redundantly) found in the valid compounds.
}
\end{table}

\subsection{Employment of the quantum annealer for sampling latent
variables from the
prior}\label{employment-of-the-quantum-annealer-for-sampling-latent-variables-from-the-prior}

The major challenge in training BM is sampling from the model
distribution. The Quantum Boltzmann Machine (QBM)
\cite{amin_quantum_2018} is the learning scheme of the quantum
Boltzmann distribution derived from a transverse field Ising Hamiltonian
using the quantum annealing processor (see
Equation~\ref{eq-qa-hamiltonian}). To the effect of QBM, we trained the
generative model with QBM or classical BM (see Methods). The loss of the
QBM model converged within 300 epochs as the classical BM model did
(Figure S3). As a consequence, the validity of compounds generated by
QBM was higher than those from classical BM (97\% in QBM and 73\% in
classical BM) (Table~\ref{tbl-metrics}).

\begin{figure}[tb]
\centering
\includegraphics[width=\linewidth]{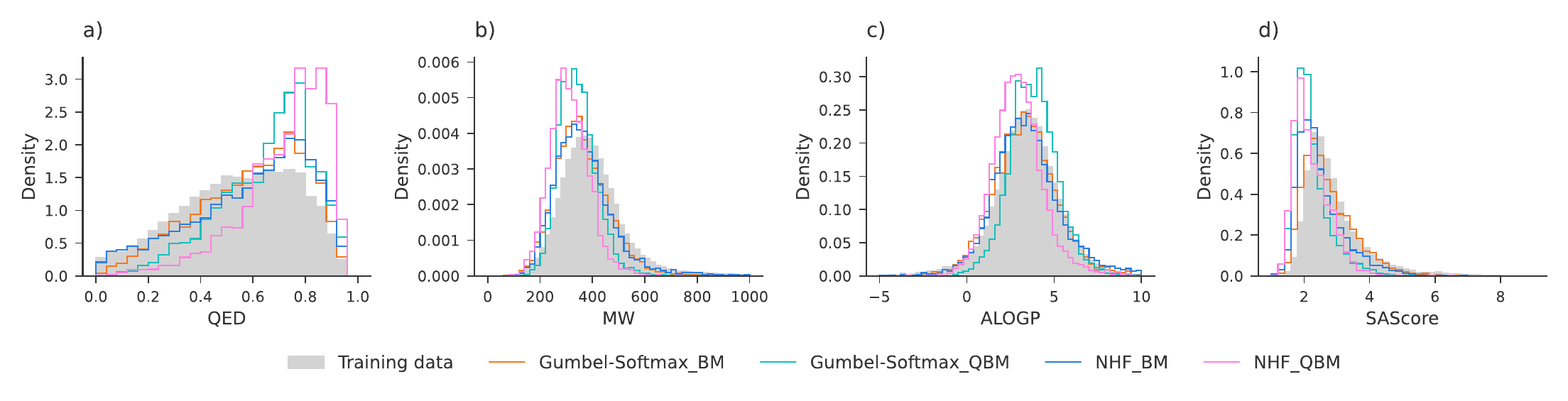}
\caption{\textbf{Distributions of various molecular
properties of generated compounds.} Three properties are shown:
drug-likeness score (QED), molecular weight (MW), lipophilicity (ALOGP),
and synthetic accessibility score (SAScore). All of these properties
were calculated by employing RDKit. Gray bars denote the training
dataset, and orange, cyan, blue and pink lines denote the samples from
the Gumbel-Softmax with classical BM, the Gumbel-Softmax with QBM, the
NHF with classical BM, and the NHF with QBM, respectively.}\label{fig-qed}
\end{figure}

Next, we compared the distributions of molecular properties and
drug-likeness score (QED score \cite{bickerton_quantifying_2012}) from
generated compounds. Surprisingly, compounds sampled from QBM shifted
their distribution of the QED score towards the higher QED side.
Moreover, the proportion of drug-like compounds (QED \textgreater{} 0.7)
was higher than those from classical BM and even the training data
(Table~\ref{tbl-drug-likeness}). This phenomenon was specific to QBM,
while samples from classical BM simply reproduced the training data
(Figure~\ref{fig-qed}). It should be noted herein that the
afore-mentioned, molecular generation beyond the training dataset in the
quality related to the drug-likeness score was achieved without any
inductions of the trends, such as some restraints and stochastic
conditions.

\begin{table}[h]
\caption{Drug-likeness of generated compounds by Transformer-based encoder-decoder models.}\label{tbl-drug-likeness}

\begin{tabular}{@{}cccc@{}}
\toprule
Data & Sampler & Binarization & Drug-like compounds (\%)
\\
\midrule
Training data     &              &                & 31.61 \\
Generated samples & Classical BM & Gumbel-Softmax & 40.81 \\
                  &              & NHF            & 43.15 \\
                  & QBM          & Gumbel-Softmax & \textbf{52.60} \\
                  &              & NHF            & \underline{\textbf{66.79}} \\
\botrule
\end{tabular}
\footnotetext{
    We defined the drug-like compounds as those with the QED score of 0.7 or higher and showed the drug-like compounds as a proportion of unique compounds.
}
\end{table}

\begin{figure}[ht]
\centering
\includegraphics[width=\linewidth]{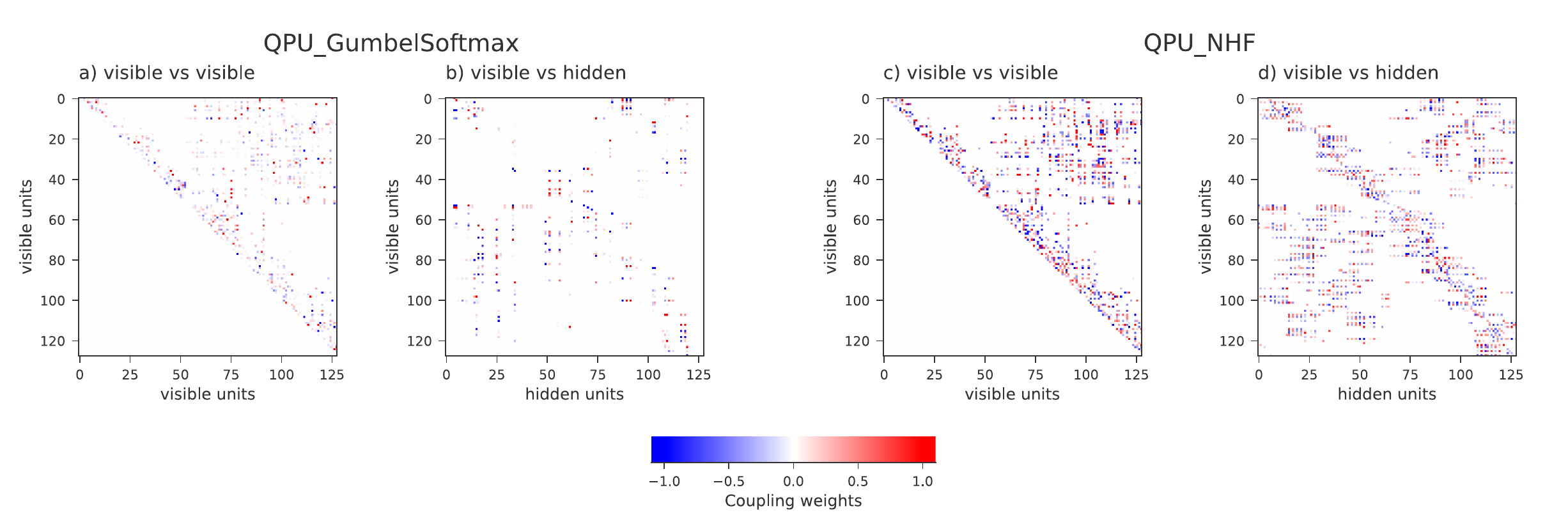}
\caption{\textbf{Coupling weights of the Ising Hamiltonian
trained in the generative models.}
(a)-(b) Heatmaps of the coupling weights between visible and visible units (\(\{J_{ij}\}\) in Equation~\ref{eq-srbm}; a) and between visible and hidden units (\(\{L_{ik}\}\) in Equation~\ref{eq-srbm}; b) of Gumbel-Softmax-based model with QPU. Note that J is the upper-triangular matrix and there are no interactions between hidden units. (c)-(d) Heatmaps of the coupling weights of NHF-based model with QPU.}\label{fig-J}
\end{figure}

Notably, under the QBM prior, the NHF-based model showed superior
validity (Table~\ref{tbl-metrics}) and drug-likeness
(Figure~\ref{fig-qed}, Table~\ref{tbl-drug-likeness}) compared to the
Gumbel-Softmax-based model. Through the training steps, the coupling
weights between the latent variables (spins) (see
Equation~\ref{eq-qa-hamiltonian}) are optimized along with the other
parameters. At the end of the training, the NHF-based model acquired
denser spin-spin coupling than the Gumbel-Softmax-based model
(Figure~\ref{fig-J}).

\begin{figure}[ht]
\centering
\includegraphics[width=\linewidth]{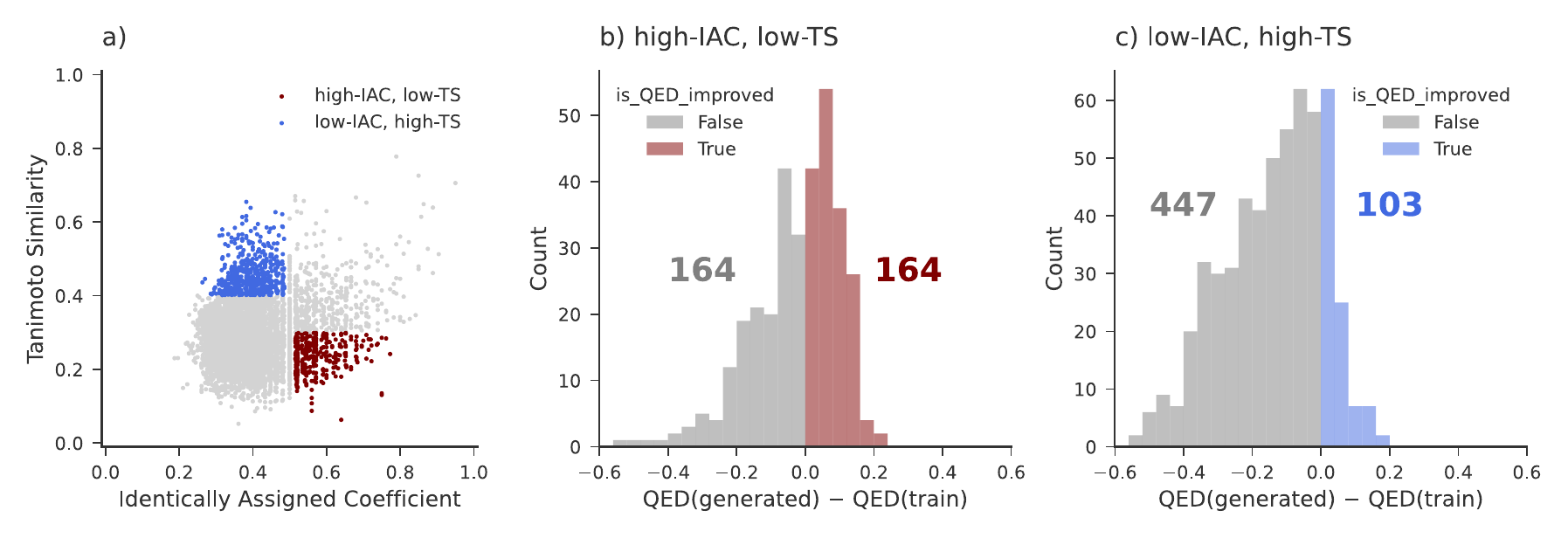}
\caption{\textbf{Relationships between structural
transformation and QED improvements via QBM models.} (a) The scatter
plot of Identically Assigned Coefficient (IAC) and Tanimoto Similarity
(TS) with respect to the subset of generated compounds, that was
selected from the QBM model so as for substructures of generated
compounds to be matched with those of the training data, as described in
the main text (see Results and Method sections). The red and blue points
denote the regions with high IAC and low TS (IAC \textgreater{} 0.5 and
TS \textless{} 0.2), and low IAC and high TS (IAC \textless{} 0.5 and TS
\textgreater{} 0.4), respectively (see Methods for detail). (b)-(c)
Differences of the QED score between the generated and original
(matched) compounds in the generated and training data, respectively. In
the subset of compounds, the features are defined by
``is\_QED\_improved'', which means that the difference of the QED score
is larger than zero. The numbers in the histograms denote those of
compounds that exhibit improved QED score values (red in (b) or blue in
(c)) and not improved QED (gray), respectively.}\label{fig-structure}
\end{figure}

Subsequently, we compared relationships between transformation of
chemical structures and improvement of the QED scores in the training
data and samples generated by the QBM model (\emph{i.e.}, the latter is
referred to herein as the generated data). We picked up 42
representative compounds from the training data and collected a subset
of compounds in the generated data, for which substructures were matched
with those of the representative compounds in the training data. This
subset of the generated data is further distinguishably comprised of
characteristic subsets using the following two similarity metrics: the
Identically Assigned Coefficient (IAC) of matched heavy atoms and the
Tanimoto Similarity (TS) of the molecular fingerprints (see Methods for
details) (Figure~\ref{fig-structure}).

More specifically, in the scatter plot of the IAC and TS
(Figure~\ref{fig-structure}), the region with high IAC and low TS values
(IAC \textgreater{} 0.5 and TS \textless{} 0.2) includes the following
compounds: the chemical structures significantly differ from the
matched, original compounds in the training data, while the molecular
sizes are comparable to them. For example, a chain group of the original
compounds in the training data is converted into ring structures, which
may thus lead to a type of ``scaffold hopping'' from the original
compounds. Conversely, the region with low IAC and high TS values (IAC
\textless{} 0.5 and TS \textgreater{} 0.4) contains compounds with their
scaffold conserved as that of the matched, original compounds in the
training data (``scaffold preserving''), although their side chains are
modified. We calculated the difference of the QED score between the
generated and original (matched) compounds in the generated and training
data, respectively. As a consequence, we found that the high-IAC and
low-TS subset (the ``scaffold hopping'' subset) exhibited a higher
proportion of compounds with improved QED compared to the low-IAC and
high-TS subset (the ``scaffold preserving'' subset) (see
Figure~\ref{fig-structure}).

The frequencies at which some substructures are found differ between
generated data and training data; for example, ether substructures are
found in training data at a frequency of \ensuremath{\sim}50\%; whereas in
generated data such substructures are found at a frequency of
\ensuremath{\sim}26\%. Since ether substructures exhibit some trends in
terms of molecular properties (e.g., ether may improve the solubility of
some compounds), we examined whether ether substructures could affect
the QED-improvement distributions by comparing the QED improvement rates
of compounds with ether substructures (see Figure~\ref{fig-structure})
versus those compounds without ether substructures (see Figure S4). The
analysis showed that compounds with ether substructures versus those
compounds without ether substructures only marginally affected the
QED-improvement distribution for generated data. In fact, the
distribution of QED improvement was almost identical for those compounds
with ether substructures as for those without, suggesting that the shift
in the QED score was not caused by any substructure bias in the
molecules generated by the QBM model. This feature of our quantum
computational system with the QBM prior and NHF-based loss function is
very useful for drug design, since scaffold hopping is still actually an
intractable task.

\subsection{Adaptation of the NHF to an MLP-based
architecture}\label{adaptation-of-the-nhf-to-an-mlp-based-architecture}

To evaluate the scope of applicability of the NHF, we additionally
performed the compound generation employing an MLP-based encoder-decoder
architecture. Using a 3-layer encoder and decoder, the validity of the
MLP-based model with the NHF (the NHF validity) was also superior to
that of Gumbel-Softmax (the Gumbel-Softmax validity)
(Table~\ref{tbl-mlp}). Compared with the afore-mentioned,
Transformer-based architecture, MLP-based architecture increased the
uniqueness of generated molecules. Thus, with a modest reduction in
validity, the number of unique compounds is comparable to that of the
Transformer-based architecture. Note that the Gumbel-Softmax validity
coupled with the QBM model was as small as 38.0\%, which means that the
MLP architecture does not work without employing NHF as a molecular
generation model.

\begin{table}[h]
\caption{Metrics of compounds generated by MLP-based encoder-decoder models.}\label{tbl-mlp}
\begin{tabular}{@{}ccccc@{}}
\toprule
Sampler & Binarization & Validity (\%) & Uniqueness (\%) &
\begin{tabular}{c}
Unique compounds / \\ generated compounds
\end{tabular}
\\
\midrule
Classical BM & Gumbel-Softmax & 38.50                      & \underline{\textbf{100.00}} & 3848 / 10000 \\
             & NHF            & \textbf{58.30}             & 74.90                       & \textbf{4367 / 10000} \\
QBM          & Gumbel-Softmax & 39.30                      & \underline{\textbf{100.00}} & 3929 / 10000 \\
             & NHF            & \underline{\textbf{59.90}} & \textbf{86.30}              & \underline{\textbf{5198 / 10000}}\\
\botrule
\end{tabular}
\footnotetext{
    Definitions of validity and uniqueness are identical to those of Table 1.
}
\end{table}

\bigskip

\section{Discussion}\label{discussion}

\subsection{Role of the NHF in our objective functions (I): beyond VAE
in posterior
approximation}\label{role-of-the-nhf-in-our-objective-functions-i-beyond-vae-in-posterior-approximation}

We developed a novel training scheme for autoencoder-based generative
models in which the latent variables are converted from continuous to
discrete by the NHF for integration with quantum annealing BM, as the
prior. As a consequence, the performance related to the molecular
generation definitively increased with respect to the two distinct types
of the autoencoder architecture (\emph{i.e.}, Transformer- and MLP-based
autoencoders) which suggests that our NHF is a robust scheme applicable
to a wide range of architectures.

Compared to the loss function of DVAE (\emph{i.e.}, \(L_\mathrm{elbo}\)
in Equation~\ref{eq-train-obj}), the KL-divergence term is replaced with
the binarization loss in the NHF (Equation~\ref{eq-loss-nhf}), which
thereby minimizes the gap between the binary and original (continuous)
latent variable vectors. In this manner, a novel loss function is
adopted in our present generative models. Although in DVAE,
KL-divergence works as the regularization term, which makes the
posterior distribution closer to the prior distribution (\emph{i.e.},
posterior approximation), balancing both reconstruction error and
regularization is difficult due to repulsive effects on the entire
objective. Most proposed methods for alleviating the shortcomings of the
conventional VAE scheme derived from KL-divergence
\cite{higgins_beta-vae_2017, tolstikhin_wasserstein_2017, zhao_infovae_2017}
require careful tuning of hyperparameters. In contrast, our presented
approach is simpler: remove the KL-divergence term and establish the
binarization of the latent variables, resulting in a constraint for the
generative AI system (\emph{i.e.}, requiring the conversion of discrete
and continuous latent variables).

Thus, let us consider the meaning of our NHF loss function from a
probabilistic perspective. We assume that the continuous vectors
\(\mathbf{h} = f_\phi(\mathbf{x}) \in \mathbb{R}^d\) as outputs from the
encoder \(f_\phi\) as well as the binary vectors
\(\mathbf{z} \in \{0,1\}^d\) are stochastic variables. We start from a
minimization of the KL-divergence between an ``approximated'' joint
posterior \(q_\phi(\mathbf{z}, \mathbf{h} | \mathbf{x})\) and a ``true''
posterior \(p_\theta(\mathbf{z}, \mathbf{h} | \mathbf{x})\)
\cite{kingma_auto-encoding_2013}, of which the latter is postulated to
involve the ``true'' distribution of data, as follows:
\begin{align}
D_\mathrm{KL} \left(
    q_\phi(\mathbf{z}, \mathbf{h} | \mathbf{x})
    \ \middle\| \ 
    p_\theta(\mathbf{z}, \mathbf{h} | \mathbf{x})
\right)
&=
\mathbb{E}_{q_\phi(\mathbf{z}, \mathbf{h} | \mathbf{x})}
\left[
    \log \frac{
        q_\phi(\mathbf{z}, \mathbf{h} | \mathbf{x})
    }{
        p_\theta(\mathbf{x}, \mathbf{z}, \mathbf{h})
    }
\right] +
\log p_\theta(\mathbf{x})
\label{eq-kl} \\
\log p_\theta(\mathbf{x}) -
D_\mathrm{KL} \left(
    q_\phi(\mathbf{z}, \mathbf{h} | \mathbf{x})
    \ \middle\| \ 
    p_\theta(\mathbf{z}, \mathbf{h} | \mathbf{x})
\right)
&=
\mathbb{E}_{q_\phi(\mathbf{z}, \mathbf{h} | \mathbf{x})}
\left[
    \log \frac{
        p_\theta(\mathbf{x}, \mathbf{z}, \mathbf{h})
    }{
        q_\phi(\mathbf{z}, \mathbf{h} | \mathbf{x})
    }
\right] \coloneq \mathcal{L}
\label{eq-kl-elbo}
\end{align}
where \(\mathcal{L}\) is the ELBO to be maximized. Next, we postulate
that the joint probability
\(p_\theta(\mathbf{x}, \mathbf{z}, \mathbf{h})\) and the approximated
posterior \(q_\phi(\mathbf{z}, \mathbf{h} | \mathbf{x})\) in
Equation~\ref{eq-kl-elbo} are factorized as follows:
\begin{align}
p_\theta(\mathbf{x}, \mathbf{z}, \mathbf{h})
&=
p_\theta(\mathbf{x} | \mathbf{z})
p_\psi(\mathbf{z})
p(\mathbf{h} | \mathbf{z})
\label{eq-factorized-joint} \\
q_\phi(\mathbf{z}, \mathbf{h} | \mathbf{x})
&=
q(\mathbf{z} | \mathbf{h})
q_\phi(\mathbf{h} | \mathbf{x})
= q_\phi(\mathbf{z} | \mathbf{x}).
\label{eq-factorized-conditional}
\end{align}

From Equation~\ref{eq-factorized-joint}, the following
holds: \(p_\theta(\mathbf{x} | \mathbf{z})\) is the likelihood of
decoder outputs
\(\hat{\mathbf{x}} = \mathrm{softmax} \left( g_\theta(\mathbf{z}) \right)\),
which is modeled by a categorical distribution
\(p_\theta(\mathbf{x} | \mathbf{z}) = \mathrm{Cat} (\mathbf{x} | \hat{\mathbf{x}})\).
\(p_\psi(\mathbf{z})\) is the prior distribution of \(\mathbf{z}\),
which is modeled by the Boltzmann Machine.
\(p(\mathbf{h} | \mathbf{z})\) is the prior distribution of
\(\mathbf{h}\). From Equation~\ref{eq-factorized-conditional}, the approximated posterior is expressed by
a single term \(q_\phi(\mathbf{z} | \mathbf{x})\) because the encoder
\(f_\phi: \mathbf{x} \mapsto \mathbf{h}\) is a ``deterministic''
function.

We define \(p(\mathbf{h} | \mathbf{z})\) in Equation~\ref{eq-factorized-joint} and
\(q_\phi(\mathbf{z} | \mathbf{x})\) in Equation~\ref{eq-factorized-conditional} as
follows:
\begin{align}
p(\mathbf{h} | \mathbf{z})
&\coloneq
\frac{1}{\sqrt{(2 \pi)^d}}
\exp
\left(
    -\frac{\lambda_\mathrm{fro}}{2}
    \|\mathbf{h} - \mathbf{z}\|^2
\right)
\label{eq-define-h-given-z} \\
q_\phi(\mathbf{z}, \mathbf{h} | \mathbf{x})
&\coloneq
\prod_{i=1}^d
q(z^{(i)} | h^{(i)}),
\hspace{0.2in}
q(z^{(i)} = 1 | h^{(i)})
\coloneq
\begin{cases}
1 & h^{(i)} \geqq 0 \\
0 & h^{(i)} < 0 \\
\end{cases}
\ \ ,
\label{eq-define-zh-given-x}
\end{align}
where \(z^{(i)} \in \{0,1\}\) and \(h^{(i)} \in \mathbb{R}\) are
\(i\)-th element of \(\mathbf{z}\) and \(\mathbf{h}\), respectively. We
maximize the expectation of the ELBO
\(\mathbb{E}_{\mathcal{D}(\mathbf{x})} \mathcal{L}\), where
\(\mathcal{D}(\mathbf{x}) \coloneq \frac{1}{N} \sum_{n=1}^N \delta(\mathbf{x} - \mathbf{x}_n)\)
denotes the empirical distribution represented by employing the delta
functions:
\begin{align}
\mathbb{E}_{\mathcal{D}(\mathbf{x})}
\mathcal{L}
&=
\mathbb{E}_{\mathcal{D}(\mathbf{x})}
\mathbb{E}_{q_\phi(\mathbf{z}, \mathbf{h} | \mathbf{x})}
\left[
    \log \frac{
        p_\theta(\mathbf{x}, \mathbf{z}, \mathbf{h})
    }{
        q_\phi(\mathbf{z}, \mathbf{h} | \mathbf{x})
    }
\right]
\nonumber \\
&=
\mathbb{E}_{\mathcal{D}(\mathbf{x})}
\mathbb{E}_{q_\phi(\mathbf{z}, \mathbf{h} | \mathbf{x})}
[\log p_\theta(\mathbf{x} | \mathbf{z})] +
\mathbb{E}_{q_\phi(\mathbf{z}, \mathbf{h})}
[\log p_\psi(\mathbf{z})]
\nonumber \\
&\ +
\mathbb{E}_{q_\phi(\mathbf{z}, \mathbf{h})}
[\log p(\mathbf{h} | \mathbf{z})] -
\mathbb{E}_{\mathcal{D}(\mathbf{x})}
\mathbb{E}_{q_\phi(\mathbf{z}, \mathbf{h} | \mathbf{x})}
[\log q_\phi(\mathbf{z}, \mathbf{h} | \mathbf{x})].
\label{eq-empirical-l}
\end{align}

Notably,
\(q_\phi(\mathbf{z}, \mathbf{h}) \coloneq \frac{1}{N} \sum_{n=1}^N q_\phi(\mathbf{z}, \mathbf{h} | \mathbf{x}_n)\)
is identical to the ``aggregated posterior'' proposed in previous
studies \cite{makhzani_adversarial_2015, hoffman_elbo_2016}. Although
Equation~\ref{eq-empirical-l} is the decomposition of the ELBO similar
to conventional VAEs
\cite{kingma_auto-encoding_2013, rezende_stochastic_2014}, we
introduced the continuous value \(\mathbf{h}\) and the binary value
\(\mathbf{z}\) as latent variables, thereby accounting for these latent
variables in our formulation. The first and second terms of
Equation~\ref{eq-empirical-l} are rewritten as follows:
\begin{align}
\mathbb{E}_{\mathcal{D}(\mathbf{x})}
\mathbb{E}_{q_\phi(\mathbf{z}, \mathbf{h} | \mathbf{x})}
[\log p_\theta(\mathbf{x} | \mathbf{z})]
&=
\frac{1}{N} \sum_{n=1}^N
\log p_\theta(\mathbf{x}_n | \mathbf{z}_n)
\nonumber \\
&=
\frac{1}{N} \sum_{n=1}^N
\log \left(
    \mathrm{Cat} (\mathbf{x}_n | \hat{\mathbf{x}}_n)
\right)
= -
\frac{1}{N} \sum_{n=1}^N
\mathrm{CE} (\mathbf{x}_n | \hat{\mathbf{x}}_n)
\label{eq-empirical-p-theta} \\
\mathbb{E}_{q_\phi(\mathbf{z}, \mathbf{h})}
[\log p_\psi(\mathbf{z})]
&=
\frac{1}{N} \sum_{n=1}^N
p_\psi(\mathbf{z}_n).
\label{eq-empirical-p-psi}
\end{align}

The third and fourth terms of Equation~\ref{eq-empirical-l} are
rewritten by using Equations~\ref{eq-define-h-given-z} and \ref{eq-define-zh-given-x} as follows:
\begin{align}
\mathbb{E}_{q_\phi(\mathbf{z}, \mathbf{h})}
[\log p(\mathbf{h} | \mathbf{z})]
&=
-\frac{\lambda_\mathrm{fro}}{2N}
\frac{1}{N} \sum_{n=1}^N
\|\mathbf{h} - \mathbf{z}\|^2
+ \mathrm{const.}
\label{eq-empirical-h-given-z} \\
\mathbb{E}_{\mathcal{D}(\mathbf{x})}
\mathbb{E}_{q_\phi(\mathbf{z}, \mathbf{h} | \mathbf{x})}
[\log q_\phi(\mathbf{z}, \mathbf{h} | \mathbf{x})]
&=
\frac{1}{N} \sum_{n=1}^N
\sum_{\mathbf{z} \in \{0, 1\}^d}
q_\phi(\mathbf{z}, \mathbf{h} | \mathbf{x}_n)
\log q_\phi(\mathbf{z}, \mathbf{h} | \mathbf{x}_n)
\nonumber \\
&=
\frac{1}{N} \sum_{n=1}^N
\sum_{i=1}^d
\sum_{\mathbf{z} \in \{0, 1\}^d}
q(z_n^{(i)} | h_n^{(i)})
\log q(z_n^{(i)} | h_n^{(i)})
= 0.
\label{eq-empirical-zh-given-x}
\end{align}

Taken together, we obtain the following equation:
\begin{align}
\mathbb{E}_{\mathcal{D}(\mathbf{x})}
\mathcal{L}
&=
-\frac{1}{N} \sum_{n=1}^N
\mathrm{CE} (\mathbf{x}_n | \hat{\mathbf{x}}_n) +
\frac{1}{N} \sum_{n=1}^N
p_\psi(\mathbf{z}_n) -
\frac{\lambda_\mathrm{fro}}{2N}
\sum_{n=1}^N
\|\mathbf{h} - \mathbf{z}\|^2
+ \mathrm{const.}
\nonumber \\
&=
-\frac{1}{N} \sum_{n=1}^N
\mathrm{CE} (\mathbf{x}_n | \hat{\mathbf{x}}_n) +
\frac{1}{N} \sum_{n=1}^N
p_\psi(\mathbf{z}_n) -
\frac{\lambda_\mathrm{fro}}{2N}
\|\mathbf{Z} - \mathbf{H}\|_F^2
+ \mathrm{const.}
\label{eq-empirical-l-detail}
\end{align}

As a result, the negation of Equation~\ref{eq-empirical-l-detail}
fundamentally matches the NHF loss (Equation~\ref{eq-loss-nhf}),
although the orthogonalization term is absent. Still, if we consider the
weight matrices of the MLP layers in the encoder
\(\mathbf{W} = \{\mathbf{W}_l\}\) as stochastic variables, the
appropriate prior distribution term \(p(\mathbf{W})\) is multiplied with
the joint prior (Equation~\ref{eq-factorized-joint}), providing the
orthogonalization constraint \cite{duan_bayesian_2020}. Thus,
Equation~\ref{eq-empirical-l-detail} exactly matches our NHF loss
function (Equation~\ref{eq-loss-nhf}).

It should be noted that the Frobenius norm term
(Equation~\ref{eq-loss-nhf}) corresponds with the cross-entropy between
\(q_\phi(\mathbf{z}, \mathbf{h})\) and
\(p_\theta(\mathbf{z}, \mathbf{h}): \mathbb{E}_{q_\phi(\mathbf{z}, \mathbf{h})}[\log p_\theta(\mathbf{z}, \mathbf{h})]\)
(Equation~\ref{eq-empirical-l-detail}). In other words, the minimization
of the Frobenius norm term forces the aggregated posterior to match the
prior, and vice versa. Since conventional VAE models the posterior
distribution using the mean-field approximation, that is,
\(q_\phi(\mathbf{z} | \mathbf{x}) = \prod_{n=1}^N q_\phi(\mathbf{z} | \mathbf{x}_n)\)
in Equation~\ref{eq-elbo}, the objective (in particular, the
KL-divergence term) independently forces each component
\(q_\phi(\mathbf{z} | \mathbf{x}_n)\) closer to the prior
\(p_\psi(\mathbf{z})\). Such a conventional protocol often leads to low
expressiveness of the latent variables (\emph{i.e.}, posterior collapse)
\cite{bowman_generating_2016, serban_hierarchical_2017}.

Moreover, for the implementation of the aggregated posterior, previous
reports introduced other architecture types, such as adversarial
networks \cite{makhzani_adversarial_2015}, which increase the
complexity of the system, requiring careful control of the training
procedure. In contrast, our NHF scheme requires only a natural
assumption on the joint prior distribution, based on the Euclidean
distance between the continuous input \(\mathbf{h}\) and the binary code
\(\mathbf{z}\) (Equation~\ref{eq-define-h-given-z}), which is equivalent
to the similarity preserving property in deep hashing
(Equation~\ref{eq-loss-nhf}). Thus, the deep hashing and aggregated
posterior schemes are unified in our NHF scheme.

Furthermore, our NHF scheme leads to the collection of the latent
variables through the Frobenius norm in the objective function (see the
following section), which corresponds to a feature driven by an
empirical distribution represented by delta functions in the aggregated
posterior scheme (Equation~\ref{eq-empirical-l-detail})
\cite{makhzani_adversarial_2015, hoffman_elbo_2016}. This
characteristic of our NHF scheme enables us to use a simpler and more
robust technique for deriving the aggregated posterior scheme as a
combination of the posterior
\(\sum_{n=1}^N q_\phi(\mathbf{z} | \mathbf{x}_n)\) and the prior.

\subsection{Role of the NHF in our objective functions (II): tractable
differentiability}\label{role-of-the-nhf-in-our-objective-functions-ii-tractable-differentiability}

A further advantage of our NHF is found in a crucial issue for AI
systems, which is related to the differentiability (\emph{i.e.},
backpropagation). In our NHF scheme, the binarization loss is defined by
the matrix norm for all samples in the minibatch
(Equation~\ref{eq-loss-nhf} and Equation~\ref{eq-empirical-l-detail}),
which is effective for approximating the gradient through the
binarization function (Equations~\ref{eq-deriv-nhf-phi} and \ref{eq-define-delta}). This is consistent
with the previous report in which the empirical loss gains smoothness
with a larger sample size, when the gradients are approximated by the
Straight Through Estimator (STE) \cite{yin_understanding_2019}. In this
way, handling the collection of the latent variables not only directly
makes the aggregated posterior and the prior closer but also reduces the
deviation of the gradient with respect to the objective function, which
is expected to guide the optimization in a more appropriate direction in
the state space of the system defined by the objective function.

\subsection{Role of the NHF in our objective functions (III): free from
reparameterization in
VAE}\label{role-of-the-nhf-in-our-objective-functions-iii-free-from-reparameterization-in-vae}

Comparing binarization schemes reveals another advantage of the NHF. In
the typical VAE, the model distribution of the posterior is limited to
the specific family for applying the reparameterization. To apply the
reparameterization, we search for a distribution
\(q_\phi(\mathbf{z} | \mathbf{x})\) that can be decomposed by the
differentiable transformation \(\xi_\phi(\mathbf{x}, \boldsymbol{\rho})\)
and auxiliary variable \(\boldsymbol{\rho} \sim p(\boldsymbol{\rho})\).
Herein, three approaches were suggested
\cite{kingma_auto-encoding_2013} for the posterior distribution: (1)
tractable and differentiable inverse cumulative distribution function
(CDF); (2) computation employing a ``standard'' distribution and its
shape parameters; and (3) their composition. Gaussian distribution is
the simplest example used in almost all VAE cases. In the discrete
cases, posterior distribution is modeled by a factorized Bernoulli
distribution and Gumbel-Softmax (Equation~\ref{eq-gumbel}) or the other
variations using the sigmoid function for smoothing
\cite{rolfe_discrete_2016}. However, such treatments can cause the
vanishing gradient. We showed that NHF models have denser connections
between latent variables (Figure~\ref{fig-J}), suggesting that the NHF
can pass the gradient more efficiently than smoothing-based methods. In
this manner, our NHF scheme exhibits a great advantage in the training
of the AI system through reducing the restriction of the distribution
shapes found in VAE.

We trained our models using the same dataset as a published model with
the conventional VAE \cite{gircha_hybrid_2023} in this work, and
revealed that the validity of the generated compounds was improved (54\%
and 97\% in the previous and present reports, respectively). Our NHF
exhibits an appropriate affinity with QBM, which can produce the
distribution distinguished from that produced by the classical BM
(Figure~\ref{fig-qed}). In fact, we showed that QBM models increased the
population of drug-like compounds, even when compared to training
datasets (Table~\ref{tbl-drug-likeness}). It should be noted herein that
we used only the tokens representing compound's chemical structures
(SMILES) for training and did not provide any information concerning
molecular properties for both training and generation processes of our
generative models. Nevertheless, the QBM model successfully generated
more drug-like compounds, as if the generative model would have known
the relationship between the structure and properties. Why was it
actualized?

A potential explanation is as follows: Because QBM-driven, enhanced
sampling with our NHF-based, extended objective function, which involves
novel posterior and prior representations, could provide more reasonable
probabilistic features, the QBM model may acquire the latent features
originating from a more generalized, stochastic distribution and, thus,
lead to a more appropriate distribution than the empirical distribution
represented by the training dataset. Consequently, our results shed
light on a new generative-model approach beyond the variational
inference leveraged by sampling from a quantum annealer.

\subsection{Role of the Boltzmann Machine as a prior
distribution}\label{role-of-the-boltzmann-machine-as-a-prior-distribution}

Most of the VAE models adopt a fixed distribution without learnable
parameters such as the standard Gaussian distribution as the prior. The
objective function of VAE tends to push the encoder to zero variance
(\emph{i.e.}, delta function), which thus leads to ``posterior holes''
in the latent space, thereby resulting in the low posterior probability
and high prior probability \cite{rezende_taming_2018}. In the
experiment, these holes are scattered in the wide region of the latent
space rather than partitioned, and so the reconstruction from the holes
generates bad samples \cite{rosca_distribution_2018}. One possible
solution to avoid them is to use a more flexible and trainable prior.
Actually, in some reports the mixture of Gaussian with the trainable
parameters was employed with respect to each component and coefficient
of the prior \cite{tomczak_vae_2018}. However, in most of these cases,
the factorial Gaussian was applied for each component, where no
correlation among the latent dimensions was postulated.

In this work, we adopted Quantum BM as a prior to resolve the
afore-mentioned issues, which thereby extended the following two
features: 1) BM naturally incorporates the correlation of the dimension
via the coupler weights \(\{J_{ij}\}\), and 2) the quantum annealer
accelerates the extended sampling for the training of the correlation
(Figure~\ref{fig-J}). Thus, the Quantum BM has a capability to
effectively capture features of the data into its parameters and
generates samples that improve the performance.

\subsection{Perspectives on adaptation of quantum-driven generative
models to drug
discovery}\label{perspectives-on-adaptation-of-quantum-driven-generative-models-to-drug-discovery}

Computer-aided molecular design is promising for efficiently finding
novel candidates appropriate for pharmaceutical drugs. Our hybrid
quantum-classical generative models can generate molecules that are
preferable in terms of the drug-likeness (QED score), even when compared
to training datasets. Moreover, our hybrid quantum-classical generative
models can also provide candidate molecules with significantly drastic
structural changes (\emph{e.g.}, from chain to ring) in generated
molecules, and, therefore, such models can be utilized in searching for
novel structures via scaffold hopping, as discussed earlier
(Figure~\ref{fig-structure}). In this work, we demonstrated that such
complicated structural modifications can be realized by using
sequence-based molecular representations like SMILES. Recently, many
reports about exploring representations, such as graph representations
\cite{lv_meta_2024, lv_meta-molnet_2025}, that are suitable for
molecular generation tasks4 have been published, indicating that with
further improvements in generative AI performance, other alternative
representations could be effective in the future.

In actual drug discovery, the scaffold is frequently fixed and
functional groups simply modified because scaffold hopping is still an
intractable issue, making it difficult to realize significant
improvements in molecular design. Furthermore, because few clues for
constructing better scaffolds have been identified, exploring the
scaffold is challenging. In contrast, the quantum-driven generative
models (machine/computer) shown in this work provide us with candidates
that can significantly improve various properties related to
drug-likeness via substantial modifications of the training dataset's
chemical structures. Then, human beings (researchers/persons) can
further modify the obtained molecular structures comprising a new
training dataset that can be fed back into the quantum-driven generative
models. Such a cyclical molecular-design process via computers and human
beings, referred to here as computer-human cyclical molecular design, is
a crucial and novel drug discovery process, thus leading to a dialogic
interactive cooperation of quantum-driven generative AI models and human
beings in the near future.

\subsection{Concluding remarks}\label{concluding-remarks}

We created a novel objective (loss) function that satisfies both
differentiability and binarization for our quantum generative models.
This loss function is free from both the mean-field approximation and
assumptions of a type of stochastic distributions (notably, both of
which are imposed in VAE), thus resulting in the improvement of outcomes
derived from the VAE architecture. As a result of the analysis beyond
VAE, the molecular properties of the obtained data generated by our
present quantum generative models outperform even those of the training
dataset. Thus, hybrid quantum and classical generative modeling is a
promising technique in the near future of drug discovery field.

\bigskip

\section{Methods}\label{methods}

\subsection{Training dataset}\label{training-dataset}

We used the subset of ChEMBL dataset provided by
\cite{gircha_hybrid_2023} (\url{https://zenodo.org/records/7827952}).
The compounds are divided into training, validation, and test sets by
8:1:1 (128,800 for training, 15,360 for validation, and 15,360 for test
sets). The identifiers of the split are included in Supplementary
Information. The canonical SMILES strings are obtained from the
molecular structures of the dataset.

\subsection{Autoencoder}\label{autoencoder}

We employed the encoder-decoder architecture of Transformer
\cite{vaswani_attention_2017} for featurizing and recomposing the
SMILES strings (Figure~\ref{fig-scheme}). Each token from SMILES is
embedded in a \(d_\mathrm{model}\)-dimensional vector and positional
encoding was added. The embedded tensors are fed into the Transformer
encoder layers. The outputs of the transformer blocks are flattened by
the Neural Tensor Network layer (for details, see the Neural Tensor
Network part in the Methods section) to obtain the fixed-length
\(D\)-dimension vectors and projected into the \(d_v\)-dimension by a
MLP layer with batch normalization for the latent variables, where
\(d_v\) is the number of the visible units of the Boltzmann machine (see
Section~\ref{sec-bm} part in Methods). In the decoder block, the binary
latent variables are embedded into the \(D\)-dimension continuous vector
by an MLP layer with batch normalization and transformed into
\(m\)-by-\(d_\mathrm{model}\) matrices for the key and the value of the
cross-attention layer. SMILES tokens, which form the embedded tensors,
are passed through both the self-attention layer with the subsequent
masks and the cross-attention layer as the query. Finally, SMILES
sequences are reconstructed from outputs of the decoder through the
softmax layer. In the generation phase, the latent variables sampled
from the prior distribution are fed into the Neural Tensor Network and
decoder, and the output tokens are predicted in an auto-regressive
manner.

\subsection{Metrics for the evaluation of generative
performance}\label{metrics-for-the-evaluation-of-generative-performance}

We generated the tokens from trained models and transformed them into
SMILES sequences. The validity was computed by dividing the number of
SMILES sequences that can be interpreted to the appropriate molecular
graph by the total number of SMILES generated. We employed RDKit
(\url{https://doi.org/10.5281/zenodo.591637}) for constructing molecule
structures from SMILES. The uniqueness was computed as a fraction of
compounds that are uniquely (not redundantly) found in the valid
compounds. We used RDKit to calculate the QED score, molecular weight
(MW), and lipophilicity (ALOGP); we used the RDKit Contrib directory's
SA\_Score descriptor
(\url{https://github.com/rdkit/rdkit/tree/master/Contrib/SA_Score}) to
calculate the synthetic accessibility score (SAScore)
\cite{ertl_estimation_2009}.

\subsection{Boltzmann machine as a prior}\label{sec-bm}

Whereas priors are conventionally fixed to a standard Gaussian
distribution, we parameterize the prior with Boltzmann machine
(BM)-based distribution powered by the quantum annealer. The BM
\cite{ackley_learning_1985} can learn a complex multi-modal probability
distribution and is thus an attractive approach for integrating quantum
computing into deep generative models. The probability distribution of
the BM is
\begin{equation}
\begin{split}
p_\psi(\mathbf{z})
&=
\frac{
    \exp \left(
        -\beta E_\psi (\mathbf{z})
    \right)
}{
    \mathcal{Z}_\psi
},
\hspace{0.3in}
\mathcal{Z}_\psi =
\sum_\mathbf{z}
\exp \left(
    -\beta E_\psi (\mathbf{z})
\right)
\\
E_\psi (\mathbf{z})
&=
\sum_{i < j}
J_{ij} z_i z_j +
\sum_{i=1}^d
h_i z_i
\end{split} \label{eq-boltzmann}
\end{equation}
where \(\mathbf{z} \in \{0, 1\}^d\) is the binary latent variables,
\(\beta\) is the inverse temperature parameter, and \(\mathcal{Z}_\psi\)
is the partition function. \(\psi = \{\mathbf{J}, \mathbf{h}\}\) are the
trainable parameters: \(J_{ij}\) denotes the interaction coefficients
between unit \(i\) and \(j\), and \(h_i\) denotes the bias term. The
Restricted Boltzmann Machine (RBM) is a subtype of BM, widely used
because of its efficient training \cite{hinton_fast_2006}. In the RBM,
a latent variable is divided into the visible units
\(\mathbf{v} = \{v_i\}_{i=1}^{d_v}\) and hidden units
\(\mathbf{u} = \{u_i\}_{i=1}^{d_u}\), where the interactions are limited
to between visible and hidden units.

We considered another BM formulation wherein the visible units form
interactions, similar to the Semi-restricted Boltzmann Machine
\cite{osindero_modeling_2007}. The Hamiltonian is defined as
\begin{equation}
E_\psi(\mathbf{v}, \mathbf{u}) =
\sum_{i < j}
J_{ij} v_i v_j +
\sum_{i=1}^{d_v}
\sum_{k=1}^{d_u}
L_{ik} v_i u_k +
\sum_{i=1}^{d_v}
b_i v_i +
\sum_{k=1}^{d_u}
c_k u_k.
\label{eq-srbm}
\end{equation}

Since the hidden units remain independent, the conditional distribution
of the hidden units can be calculated independently by
\begin{equation}
p(u_k | \mathbf{v}) =
\mathrm{tanh} \left(
    u_k \lambda_k(\mathbf{v})
\right),
\hspace{0.3in}
\lambda_k(\mathbf{v}) =
c_k +
\sum_{i=1}^{d_v}
L_{ik} v_i.
\label{eq-rbm-cond}
\end{equation}

We applied the distribution marginalized over the hidden units
\begin{equation}
p_\psi(\mathbf{v}) =
\frac{
    \sum_\mathbf{u} \left(
        -\beta E_\psi(\mathbf{v}, \mathbf{u})
    \right)
}{
    \mathcal{Z}_\psi
},
\label{eq-rbm-marginal}
\end{equation}
to the prior distribution. Therefore, the KL-divergence term of the
objective function is given by
\begin{equation}
D_\mathrm{KL} \left(
    q_\phi(\mathbf{z} | \mathbf{x})
    \ \middle\| \ 
    p_\psi(\mathbf{z})
\right) =
\mathbb{E}_{q_\phi(\mathbf{z} | \mathbf{x})}
[\log q_\phi(\mathbf{z} | \mathbf{x})] -
\mathbb{E}_{q_\phi(\mathbf{z} | \mathbf{x})}
[\log p_\psi(\mathbf{z})],
\label{eq-rbm-kl}
\end{equation}
and the gradient of the second term is calculated by the difference of
energy between samples from the data (positive phase) and from the prior
(negative phase) as
\begin{align}
\mathbb{E}_{q_\phi(\mathbf{z} | \mathbf{x})}
\left[
    \partial \log p_\psi(\mathbf{z})
\right]
&=
\mathbb{E}_{\mathbf{z} \sim q_\phi(\mathbf{z} | \mathbf{x})}
[\partial F_\psi(\mathbf{z})] -
\mathbb{E}_{\mathbf{v}, \mathbf{u} \sim p_\psi(\mathbf{v}, \mathbf{u})}
[\partial E_\psi(\mathbf{v}, \mathbf{u})]
\label{eq-rbm-kl-deriv} \\
F_\psi(\mathbf{z})
&=
\sum_{i < j}
J_{ij} z_i z_j +
\sum_{i=1}^{d_v}
b_i z_i -
\sum_{k=1}^{d_u}
\log 2 \mathrm{cosh}(-\lambda_k (\mathbf{z})),
\label{eq-free-energy}
\end{align}
where \(F_\psi\) is the marginalized Hamiltonian (free energy) of
visible units. This equation means that the first term is calculated by
the average of the latent variables from the data and that the second
term is calculated using samples from the prior distribution. We used
simulated annealing with Metropolis-Hastings update to approximate
sampling from BM.

\subsection{Quantum Boltzmann machine and its
implementation}\label{quantum-boltzmann-machine-and-its-implementation}

We use the D-Wave quantum annealer as a source of prior samples. When
the annealing time is sufficiently long with quasistatic evolution, the
quantum processing unit (QPU) generates samples that approximate a
classical Boltzmann distribution \cite{amin_searching_2015}. However,
the presence of a finite transverse field introduces quantum
fluctuations that can lead to deviations from the ideal Boltzmann
statistics. A quantum Boltzmann machine (QBM) \cite{amin_quantum_2018},
which draws samples from the Boltzmann distribution of a
transverse-field Ising Hamiltonian, can be trained analogously to its
classical counterpart by optimizing a variational bound on the true
log-likelihood. In this study, we introduce an RBM architecture that is
restricted to the Zephyr graph (\emph{i.e.}, an architecture native to
the D-Wave QPU), thus allowing for more efficient sampling and larger
dimensions than a traditional architecture, such as bipartite and clique
graphs, which are not native to the Zephyr graph. In this study, the
largest possible clique and bipartite graphs that could be built on a
defect-free Zephyr architecture are \(K_{2(2m-1)t}\) and
\(K_{2(2m-1)t,2(2m-1)t}\), respectively, where \(m\) is the grid
parameter and \(t\) is the tile parameter \cite{boothby_zephyr_2021}.
The QPU used in this study has a grid parameter of 6 and a tile
parameter of 4, only allowing a clique of 88 and a \(K_{88,88}\)
(\emph{i.e.}, a bipartite graph with 88 units on the visible side and 88
units on the hidden side).

To utilize the entire QPU as an RBM prior and allow for a flexible
number of visible and hidden units, we use chains of qubits to achieve
the desired prior dimension as well as increase the connectivity between
visible units. The choice of chains and hidden nodes is defined through
a heuristic node-contraction method where a sparse graph is iteratively
converted to a denser graph, starting with a chain length of two and
increasing the length as needed. In this method, nodes are first sorted
by degree. The node with the lowest degree along with its adjacent
nodes, which also have the smallest degree, are chosen for contraction.
When the resulting contracted graph reaches the required dimensions, the
contraction stops. The node-contraction method works with any graph and
especially with D-Wave's current and future quantum annealing
architectures. One advantage of this method is that it works with QPUs
with imperfect graphs and allows for a granular control over the number
of visible and hidden variables. In our implementation of the Boltzmann
machine, the hidden units are not connected to each other, which is
compatible with the Zephyr graph as there are groups of qubits that are
not connected to each other. Figure S1 (see Supporting Information)
illustrates the classification of qubits within the Zephyr architecture
into four distinct groups, referred to as color groups. Color groups are
suitable candidates for hidden units, because a color group does not
have any internal connections, whereas different color groups have
external connections between each other. In the Advantage2\_prototype2.6
QPU, each color group contains more than 300 qubits and all hidden units
are selected exclusively from a single-color group to ensure compliance
with the no-connectivity constraint among hidden units; that is,
selecting hidden variables from multiple color groups would violate this
constraint. If the number of hidden units were to exceed 300, couplers
would need to be removed to adhere to the no-connectivity condition
between hidden units; however, 300 hidden units are deemed sufficient
for the scope of this study. Figure S2 (see Supporting Information)
shows a Zephyr graph that we used as an RBM prior, having 128 visible
and 128 hidden units. Note that we utilized the entire QPU (1215 qubits)
in this prior, where chains were added to increase the connectivity of
visible units.

When sampling from the QPU, one may need to know the effective
temperature (\(\beta\)) at which sampling is performed. We could
estimate this temperature using the maximum log-likelihood approach
based on the model parameters and rate of excitations
\cite{raymond_global_2016}, which are available with a D-Wave utility
(\url{https://docs.dwavequantum.com/en/latest/ocean/api_ref_system/generated/dwave.system.temperatures.maximum_pseudolikelihood_temperature.html}).
In this study, we scale \(J\) and \(b\) parameters of the Hamiltonian
such that the QPU produces samples with an effective temperature near
1.0.

\subsection{Neural Tensor Network}\label{neural-tensor-network}

A compound in the training data is embedded to tensor
\(\mathbf{X} \in \mathbb{R}^{C \times d_\mathrm{model}}\), where \(C\)
is the number of token sequences and \(d_\mathrm{model}\) is the
embedding dimension. Since the sequence length varies for each compound,
the tensor size is also variable. To obtain the latent vector with fixed
length, it is necessary to convert the variable length tensor to the
fixed length one. The simplest way is to aggregate the vectors of each
length by the pooling (max, average, etc.). However, this can cause
information loss concerning the length. Another method is
convolution/deconvolution for converting to the fixed length vector,
wherein the range of the length is fixed in advance for determining the
strides. We propose a more flexible form, which is referred to here as
Neural Tensor Network (NTN), using tensor products with trainable
higher-order tensors. For the end of the encoder (NTN1)
(Figure~\ref{fig-scheme}), conversion from the variable length tensor
\(\mathbf{X}\) to fixed length vector \(\mathbf{h} \in \mathbb{R}^D\) is
performed as follows:
\begin{equation}
b_\alpha =
\sum_{i=1}^{d_\mathrm{model}}
\sum_{j=1}^{d_\mathrm{model}}
\sum_{k=1}^C
W_{ij\alpha}^{(1)}
X_{ki} X_{kj},
\hspace{0.3in}
h_\alpha = \sigma(b_\alpha),
\label{eq-ntn1}
\end{equation}
where \(\mathcal{W}^{(1)} = \{W_{ij \alpha}^{(1)}\} \in \mathbb{R}^{d_\mathrm{model} \times d_\mathrm{model} \times D}\)
is a trainable tensor and \(\sigma(\cdot)\) is an element-wise
activation function. For the beginning of the decoder (NTN2)
(Figure~\ref{fig-scheme}), conversion from the latent vector
\(\boldsymbol{\zeta} \in \mathbb{R}^D\) to the tensor
\(\mathbf{M} \in \mathbb{R}^{m \times d_\mathrm{model}}\) is performed
as follows:
\begin{equation}
B_{li} =
\sum_{\alpha=1}^D
W_{\alpha li}^{(2)} \zeta_\alpha,
\hspace{0.3in}
M_{li} = \sigma(B_{li}),
\label{eq-ntn2}
\end{equation}
where
\(\mathcal{W}^{(2)} = \{W_{\alpha li}^{(2)}\} \in \mathbb{R}^{D \times m \times d_\mathrm{model}}\)
is a trainable tensor.

\subsection{Derivation of the gradient of NHF with respect to the
encoder
parameters}\label{derivation-of-the-gradient-of-nhf-with-respect-to-the-encoder-parameters}

Using chain rules, the gradient of the loss in Neural Hash Function
\(L_\mathrm{nhf}\) (Equation~\ref{eq-loss-nhf}) with respect to \(\phi\)
is
\begin{align}
\frac{
    \partial L_\mathrm{nhf}
}{
    \partial \phi
}
&=
\frac{
    \partial L_\mathrm{nhf}
}{
    \partial \mathbf{z}_n
}
\frac{
    \partial \mathbf{z}_n
}{
    \partial \mathbf{h}_n
}
\frac{
    \partial \mathbf{h}_n
}{
    \partial \phi
}
\nonumber \\
&=
\sum_{n=1}^N
\left(
    \frac{
        \partial L_\mathrm{rec}
    }{
        \partial \mathbf{z}_n
    } +
    \frac{
        \partial L_\mathrm{prior}
    }{
        \partial \mathbf{z}_n
    } +
    \frac{
        \partial L_\mathrm{quant}
    }{
        \partial \mathbf{z}_n
    }
\right)
\frac{
    \partial \mathbf{z}_n
}{
    \partial \mathbf{h}_n
}
\frac{
    \partial \mathbf{h}_n
}{
    \partial \phi
}.
\label{eq-dlnhf-dphi}
\end{align}

Using vector-form, \(L_\mathrm{quant}\) can be rewritten as
\begin{align}
L_\mathrm{quant}
&=
\frac{
    \lambda_\mathrm{fro}
}{
    2N
}
\|\mathbf{Z} - \mathbf{H}\|_F^2 +
\frac{
    \lambda_\mathrm{ortho}
}{
    2
}
\|\mathbf{W}_l \mathbf{W}_l^T - \mathbf{I}\|_F^2
\nonumber \\
&=
\frac{
    \lambda_\mathrm{fro}
}{
    2N
}
\sum_{n=1}^N
\|\mathbf{z}_n - \mathbf{h}_n\|^2 +
\frac{
    \lambda_\mathrm{ortho}
}{
    2
}
\|\mathbf{W}_l \mathbf{W}_l^T - \mathbf{I}\|_F^2
\label{eq-lquant-vec}
\end{align}
and the gradient of \(L_\mathrm{quant}\) with respect to
\(\mathbf{z}_n\) are computed to
\begin{equation}
\frac{
    \partial L_\mathrm{quant}
}{
    \partial \mathbf{z}_n
} =
\frac{
    \lambda_\mathrm{fro}
}{
    2
}
(\mathbf{z}_n - \mathbf{h}_n).
\label{eq-dlquant-dzn}
\end{equation}

When the transformation from \(\mathbf{h}_n\) to \(\mathbf{z}_n\) is
non-smooth, such as the Sign function or Heaviside step function, the
gradient \(\frac{\partial \mathbf{z}_n}{\partial \mathbf{h}_n}\) cannot
be computed. We approximated this derivative by the identity function.
With this approximation, the backward pass is different from the forward
pass, whereas the sign of negation of the backward pass is consistent
with the direction that minimizes the loss
\cite{yin_understanding_2019, hinton_neural_2012}. From
Equation~\ref{eq-dlnhf-dphi} and Equation~\ref{eq-dlquant-dzn}, we
obtained
\begin{equation}
\frac{
    \partial L_\mathrm{nhf}
}{
    \partial \phi
} =
\sum_{n=1}^N
\left(
    \frac{
        \lambda_\mathrm{fro}
    }{
        2
    }
    (\mathbf{z}_n - \mathbf{h}_n) +
    \frac{
        \partial L_\mathrm{rec}
    }{
        \partial \mathbf{z}_n
    } +
    \frac{
        \partial L_\mathrm{prior}
    }{
        \partial \mathbf{z}_n
    }
\right)
\frac{
    \partial \mathbf{h}_n
}{
    \partial \phi
}.
\label{eq-dlnhf-dphi-approx}
\end{equation}

\subsection{Relationships between transformation of chemical structures
and improvements of QED
score}\label{relationships-between-transformation-of-chemical-structures-and-improvements-of-qed-score}

We selected the representative compounds from the training data using
the following criteria: 1) Molecular Weight (MW) greater than 200 and
less than 450, 2) logP greater than 0 and less than 4, 3) PSA less than
150, and 4) without alert structures (computed by RDKit). From the
generated compounds, we extracted those that shared the same
substructures as those of each of representative compounds. As queries
of substructure search, a pool of fragments is required to consist of
chemically meaningful components. We created fragments from the
representative compounds using RECAP
\cite{lewell_recapretrosynthetic_1998} and BRICS \cite{degen_art_2008}
fragmentation and filtered by MW (greater than 150). Herein, for a
similarity measure, the Identically Assigned Coefficient (IAC) of atoms
was computed as follows
\begin{equation}
\begin{split}
\mathrm{IAC} =
\frac{
    \mathrm{HAC}_\mathrm{matched}
}{
    \mathrm{HAC}_\mathrm{generated} +
    \mathrm{HAC}_\mathrm{train} -
    \mathrm{HAC}_\mathrm{matched}
}
\\
\end{split} \label{eq-iac}
\end{equation}
where \(\mathrm{HAC}_\mathrm{generated}, \mathrm{HAC}_\mathrm{train}\),
and \(\mathrm{HAC}_\mathrm{matched}\) denote the Heavy Atom Count (HAC)
of the generated compound, the training data, and the matched
substructure, respectively. For another similarity measure, we used the
Tanimoto Similarity of the molecular fingerprint. Molecular fingerprints
are generated as 2048-bit vectors using Morgan algorithm
\cite{morgan_generation_1965} with radius of 2, which is roughly
equivalent to ECFP4 fingerprint
\cite{rogers_extended-connectivity_2010}.

\subsection{Training hyperparameters}\label{training-hyperparameters}

As inputs of the autoencoder, the tokens of SMILES were embedded into
160-dimension vector. For the encoder and the decoder, 5 stacks of the
transformer layers with the 1024-dimension feedforward layers, the ReLU
activation functions, and the Dropout with rate of 0.1 were used. The
output dimension of the Neural Tensor Network 1 (NTN1) was 1024 and the
output of the NTN2 was a 128-by-160 matrix. We set the number of visible
and hidden units in the Boltzmann machine as 128 and 128, respectively,
and trained models with the minibatch of 512 for 300 epochs, using the
Adam optimizer \cite{kingma_adam_2014} (\(\beta_1\) = 0.9, \(\beta_2\)
= 0.999) with a learning rate of 10\textsuperscript{-4} and a weight decay of 10\textsuperscript{-3}.

\bigskip

\section{Data Availability}\label{data-availability}

In the current study, we used the subset of ChEMBL dataset provided by
the previous report \cite{gircha_hybrid_2023}
(\url{https://doi.org/10.5281/zenodo.7827952}). The code and the
generated molecules during the current study are available from the
corresponding author on reasonable request.

\bigskip

\section{References}\label{references}

\renewcommand{\bibsection}{}
\bibliography{bibliography.bib}

\bigskip

\section{Acknowledgements}\label{acknowledgements}

This study received no funding.

\bigskip

\section{Author contributions}\label{author-contributions}

H.K., M.R. Y.I., V.V.C., W.K., K.C., M.A.~and M.T. designed research.
H.K., Y.I., Y.H., A.S. and M.T. constructed the generative models, ran
experiments and analyzed the data. M.R., M.W., V.V.C., W.K., K.C. and
M.A.~contributed to implementation of the quantum annealing. All the
authors discussed the results and contributed to writing the manuscript.

\bigskip

\section{Competing interests}\label{competing-interests}

The authors have no competing interests.

\end{document}